\documentclass[10pt]{article}
\usepackage[OE]{express}
\usepackage{soul}
\setstcolor{red}

\begin{document}

\title{Preparation of entangled states of microwave photons in a hybrid system via electro-optic effect}

\author{Daoquan Zhu, Pengbo Li\authormark{*}}

\address{Department of Applied Physics,
Xi'an Jiaotong University, Xi'an 710049, China}

\email{\authormark{*}lipengbo@mail.xjtu.edu.cn} 



\begin{abstract}
We propose to realize the two-mode continuous-variable entanglement of microwave photons in an electro-optic system, consisting of two superconducting microwave resonators and one or two optical cavities filled with certain electro-optic medium. The cascaded and parallel schemes realize such entanglement via coherent control on the dynamics of the system, while the dissipative dynamical scheme utilizes the reservoir-engineering approach and exploits the optical dissipation as a useful resource. We show that, for all the schemes, the amount of entanglement is determined by the ratio of the effective coupling strengths of "beam-splitter" and "two-mode squeezing" interactions, rather than their amplitudes.
\end{abstract}

\ocis{(270.5585) Quantum information and processing; (270.6570) Squeezed states.} 

\bibliographystyle{osajnl}
\bibliography{thesis_ver8}
\section{Introduction}
With the development of quantum information, microwave radiation has become an important tool to couple different quantum systems due to its frequency range covering many types of qubits \cite{RevModPhys.85.623}. It can couple different qubits to realize quantum computation \cite{PhysRevB.68.064509,1612-202X-13-8-085201,PhysRevA.94.012328,PhysRevA.88.024303,PhysRevA.88.063632,PhysRevA.84.010301,0295-5075-85-5-50007,Feng20081589,PhysRevA.75.064303,PhysRevA.69.062320}, and deterministic logic operations \cite{PhysRevLett.100.170501,PhysRevA.71.052310}. Also people employ microwaves to trap charged particles for quantum information processing \cite{Piltze1600093,Hensinger2011Quantum,Ospelkaus2011Microwave,PhysRevA.71.012315}.

Unlike optical photons, it is hard to entangle microwave photons through nonlinear optical methods with optical crystals. It is thus appealing to propose alternative approaches to generate entangled microwave photons. Many different approaches have been explored in different systems. For instance, some theoretical works have presented the dissipation-based approach in electro-mechanical systems \cite{PhysRevA.88.043802}, the coherent-control-based approach through excitations of cavity Bogoliubov modes in optomechanical systems\cite{PhysRevLett.110.233602} and electro-mechanical systems\cite{PhysRevA.88.043802}, as well as the schemes utilizing solid-state superconducting circuits \cite{PhysRevA.90.062342,PhysRevA.87.022320,PhysRevLett.106.060401,abcd-1367-2630-19-2-023027,Cheng:13,PhysRevLett.109.250502,Lu:09}.

In previous works, it has been demonstrated that the electro-optic coupling has the same form as optomechanical\cite{Matsko:07} and electro-mechanical couplings\cite{PhysRevA.88.043802}. So all previous considered effects can in principle be observed in electro-optic systems \cite{PhysRevA.81.063837}. But there are several challenges of optomechanical and electro-mechanical systems to cool down their nano-/micro- mechanical oscillators. First, due to the low frequencies of mechanical oscillators, the required environment temperature is ultra low. Moreover, there's a limitation of physical cooling as demonstrated in \cite{RevModPhys.86.1391} that mechanical oscillators can not be cooled down to their ground states under the influence of cavity frequency noise. Besides cooling processes, the quality factors of high-frequency mechanical oscillators are relatively low. However, we can avoid those drawbacks of micro mechanical oscillators through electro-optic systems. Auxiliary modes in electro-optic systems are optical modes, which can be regard as vacuum states at experimental temperatures. In addition, with the well-developed fabrication of optical cavities and superconducting circuits, it is convenient to get optical cavities with desired quality factors as well as low-dissipation superconducting microwave resonators to prepare high-quality entanglement states.

In this work, inspired by optomechanical systems\cite{RevModPhys.86.1391,PhysRevA.89.022332,PhysRevA.83.033820,PhysRevA.92.062311,Jiang:13, PhysRevA.92.023856,PhysRevA.78.032316,PhysRevA.84.042342,PhysRevLett.103.213603,PhysRevLett.109.130503,PhysRevA.80.065801,Kippenberg1172} and other hybrid quantum systems\cite{PhysRevA.88.043802,PhysRevA.81.063837,Chen:15,Matsko:07}, we propose an electro-optic system comprising two separated superconducting microwave resonators and one or two auxiliary optical cavities, which are filled with certain electro-optic medium. With this system, we provide three schemes to entangle these two microwave resonators via electro-optic effect: (i) cascaded scheme; (ii) parallel scheme; (iii) dissipative dynamical scheme. The underlying physics for both cascaded and parallel schemes is the coherent control over their systems to realize the Bogoliubov modes consisting of the two microwave modes, while the last scheme is based on quantum reservoir engineering, which exploits the dissipations of two optical cavities as useful resources to entangle microwave photons.  For each scheme, we've worked out the analytic solutions and numerical simulations. In Section \ref{sec1}, we also talk about the experimental feasibility of all schemes. Especially, Eq.(\ref{eq43})-Eq.(\ref{eq45}) shows that the temperature dependence of the entanglement degree for the dissipative dynamical scheme is steerable. It's modulated by the decay rate of both optical cavities and superconducting microwave resonators, as well as the ratio of effective coupling strengths. Therefore, high quality entanglement can be realized through choosing cavities and resonators of optimized quality factors.

\section{The Model and schemes}
\subsection{The cascaded scheme}
\begin{figure}[!ht]
\graphicspath{{fig/}}
  \centering
  \includegraphics[width=8cm]{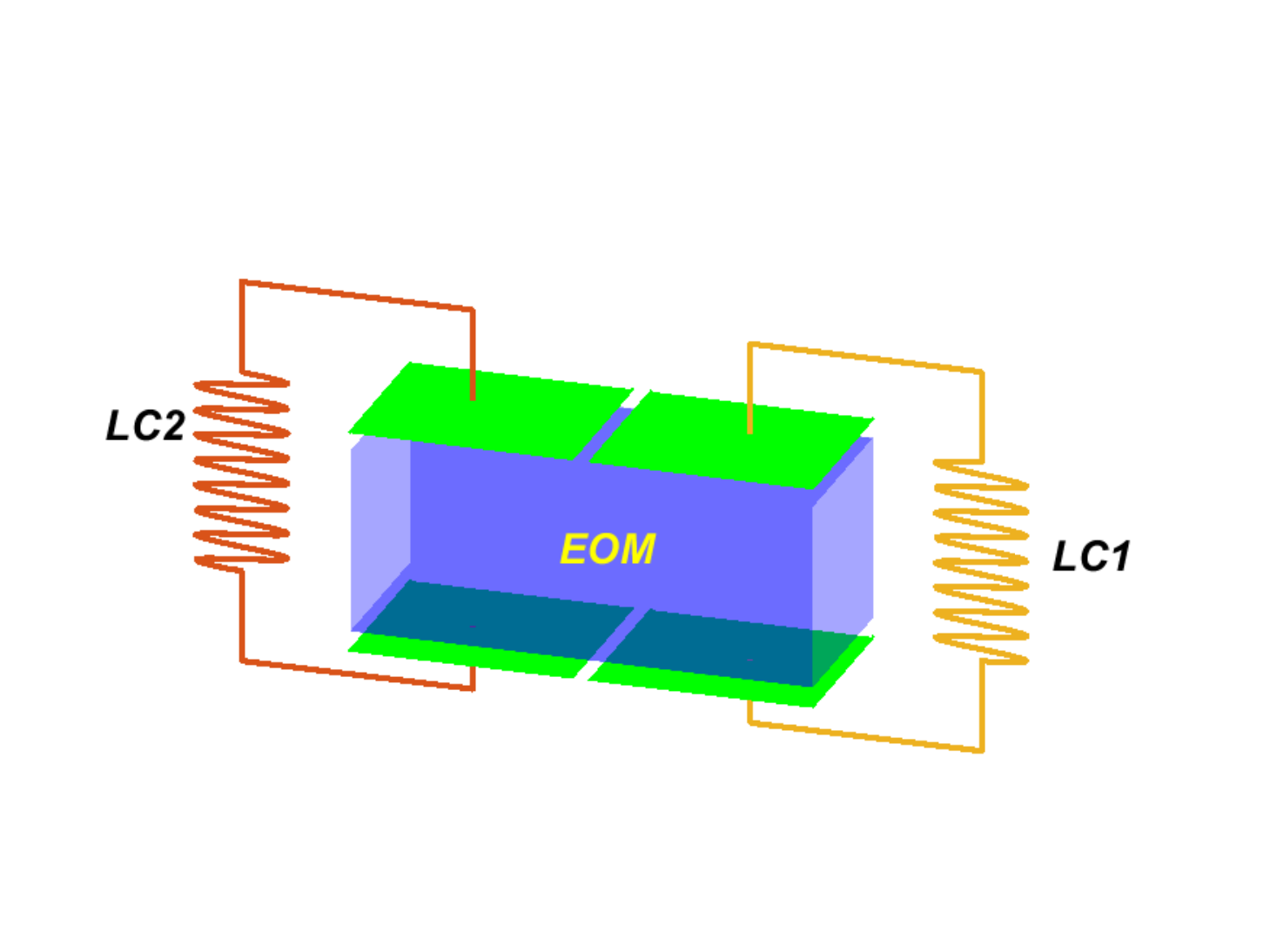}\\
  \caption{The setup of both cascaded and parallel schemes. In the cascade scheme, the optical cavity is driven by a laser of frequency $\omega_{L1}$ or $\omega_{L2}$ for different periods, while in the parallel scheme, we impose both of these driving lasers at the same time. \label{fig:1}}
\end{figure}

As shown in Fig.~\ref{fig:1}, the hybrid quantum system we considered  composes of two microwave resonators with frequency $\omega_{b1}$(LC1) and $\omega_{b2}$(LC2), and an optical cavity of frequency $\omega_{a1}$, inside which a kind of electro-optic medium(EOM) such as KDP is filled. These two resonators are coupled to the optical cavity through electro-optic effect, but have no direct interaction with each other.
 
\begin{figure}[!ht]
\graphicspath{{fig/}}
  \centering
  \includegraphics[width=12cm]{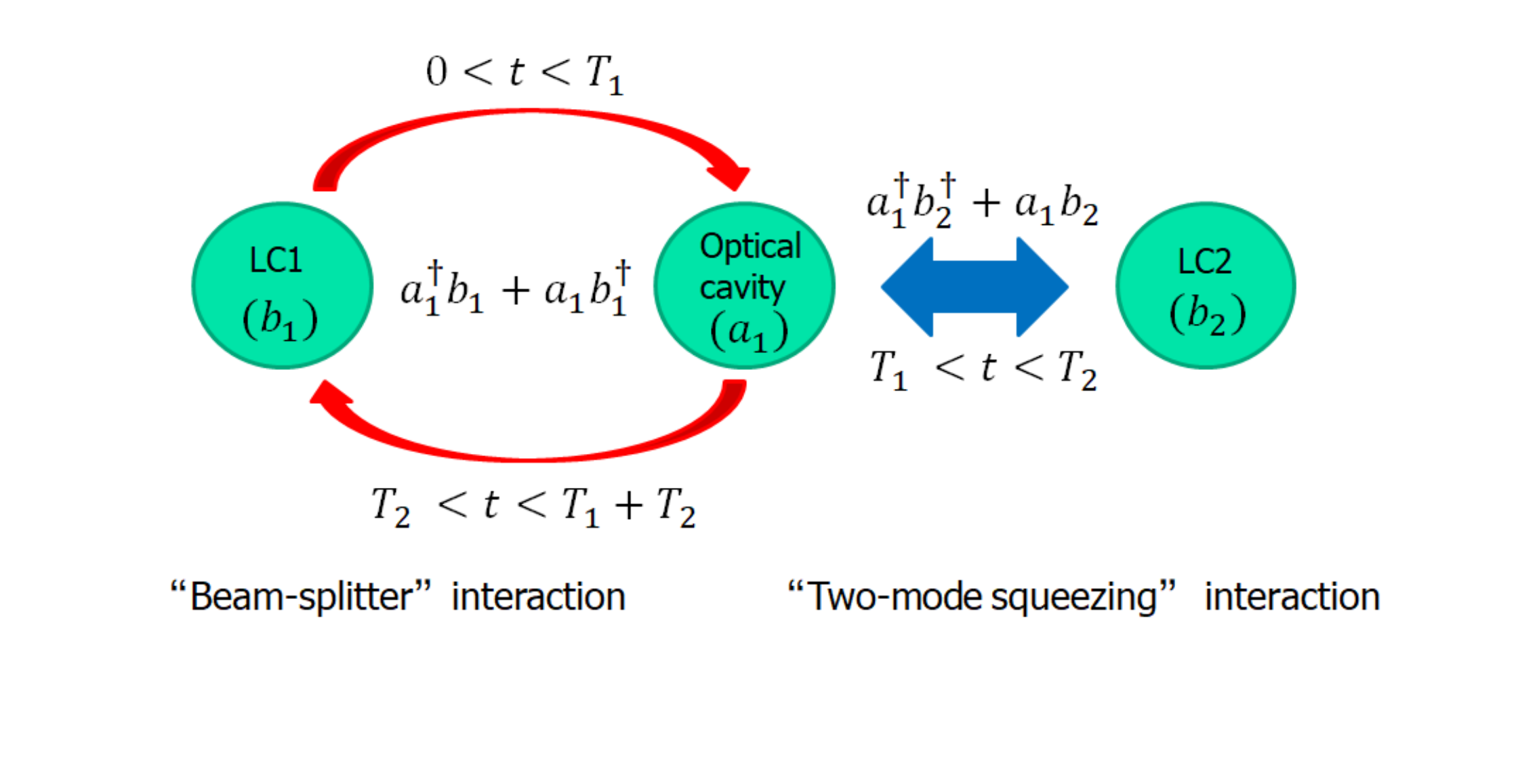}\\
  \caption{A diagram of the process of the cascaded scheme, and $a_1$, $b_1$, $b_2$ are annihilation operators of optical mode and two microwave modes, respectively. When $0<t<T_1$ or $T_2<t<T_1+T_2$, LC1 and the optical cavity exchange quantum states with each other, and during $T_1<t<T_2$, modes of optical cavity and LC2 get entangled. \label{fig:0}}
\end{figure}

It is known that the effect of "beam-splitter" interaction is to exchange quantum states between two modes, and that of "two-mode squeezing" interaction is to get both modes entangled. Therefore, one straightforward approach to entangle the microwave photons in these separated resonators is to drive the optical cavity of this system with suitably detuned lasers in a cascaded way as shown in Fig.~\ref{fig:0}: (i) to set LC1 in the red-detuned regime; (ii) to set LC2 in the blue-detuned regime; (iii) to set LC1 in the red-detuned again. Then at the final moment, the Bogoliubov modes composed of the two resonator modes only will be excited.

The detailed steps of this scheme are as the following. We drive the optical cavity with different lasers in different periods: (i) $0<t<T_1$, using the laser of frequency $\omega_{L1}$; (ii) $T_1<t<T_2$, using the laser of frequency $\omega_{L2}$; (iii) $T_2<t<T_1+T_2$, we use the laser of frequency $\omega_{L1}$, again. We assume $\omega_{L1}-\omega_{a1}=-\omega_{b1},~\omega_{L2}-\omega_{a1}=\omega_{b2}$ to guarantee that the microwave resonators are in the suitable detuned regimes. Through the rotating-wave approximation, in each period there is only one microwave mode interacting with the optical mode. In other words, we set LC1 to be in the red-detuned regime and LC2 to be in the blue-detuned regime when they interact with the optical cavity, and both of them are isolated when they are far detuned from the optical cavity.
As shown in \cite{PhysRevA.81.063837}, when we only consider one mode of the optical cavity, the interaction Hamiltonian in each period is:
\begin{equation}
H_{I,i}^C=-\hbar g_i a_1^{\dag} a_1 (b_i+b_i^{\dag}),~\label{eq4}\\
\end{equation}
\begin{equation}
g_i=\frac{\omega_{a1} n^3 r_0}{2d} \sqrt{\frac{\hbar \omega_{i}}{2C_i}},~\label{eq5}
\end{equation}
where $n$, $r_0$, and $d$ are the refractive index, electro-optic coefficient and height of the medium respectively. $C_i$ refers to the capacitance of the $i^{th}$ resonator.

In the first period, the driven term in the total Hamiltonian is that:
\begin{equation}
H_d^C=i \hbar [E_0 a_1^{\dag} e^{-i \omega_{L1} t}-E_0^* a_1 e^{i \omega_{L1} t}],
\end{equation}
where $E_0$ is the complex amplitude of the driving laser. If we choose a rotating frame with frequency $\omega_{L1}$ respect to the optical mode $a_1$, the total Hamiltonian then becomes:
\begin{gather}
H_1^C=-\hbar \Delta a_1^{\dag}a_1+\hbar \omega_{b1} b_1^{\dag}b_1+\hbar \omega_{b2} b_2^{\dag}b_2\notag\\
-\hbar g_1 a_1^{\dag}a_1(b_1^{\dag}+b_1)+i\hbar (E_0 a_1^{\dag}-E_0^*a_1),~\label{eq7}
\end{gather}
where $\Delta=\omega_{L1}-\omega_{a1}$. In Eq.(\ref{eq7}), we have assumed that the driving laser is strong enough. Therefore, it is a good approximation to linearize the above Hamiltonian through replacing the optical annihilate operator $a_1$ by the sum of its stable mean value $\bar{a}$ and its fluctuation term $\delta a$. The interaction term between the optical mode and LC2 mode has been eliminated by the rotating wave approximation. Employing Heisenberg equation, the zero order and linear terms are eliminated, so we only consider the quadratic terms. Then Eq.(\ref{eq7}) becomes:
\begin{gather}
H^C_{1,eff}=-\hbar\Delta\delta a_1^{\dag}\delta a_1+\hbar\omega_{b1} a_1^{\dag}b_1+\hbar\omega_{b2} b_2^{\dag}b_2\notag\\
-\hbar\bar{a}_1 g_1(\delta a_1^{\dag} b_1+\delta a_1 b_1^{\dag}).~\label{eq8}
\end{gather}
The effective coupling strength in the first period is $\bar{a}_1g_1$, which can also be modulated by the power of the driving lasers. For simplicity we introduce the non-dimension time: $\tau=\bar{a}_1g_1t,~\tau_i=\bar{a}_1g_1T_i,~i=1,2$. Then in the interaction picture, the Heisenberg equations for all three modes can be written as:
\begin{equation}
\frac{d}{d\tau}
\left[
\begin{array}{ccc}
\delta a_1(\tau)\\b_1(\tau)\\b_2^{\dag}(\tau)
\end{array}\right]
=\left[
\begin{array}{ccc}
0&i&0\\i&0&0\\0&0&0
\end{array}\right]
\left[\begin{array}{ccc}
\delta a_1(0)\\b_1(0)\\b_2^{\dag}(0)
\end{array}\right].
\end{equation}
The solution of these equations is straightforward:
\begin{equation}
\left[
\begin{array}{ccc}
\delta a_1(\tau)\\b_1(\tau)\\b_2^{\dag}(\tau)
\end{array}\right]
=\left[
\begin{array}{ccc}
\cos(\tau)&i\sin(\tau)&0\\i\sin(\tau)&\cos(\tau)&0\\0&0&1
\end{array}\right]
\left[\begin{array}{ccc}
\delta a_1(0)\\b_1(0)\\b_2^{\dag}(0)
\end{array}\right].
\end{equation}
Similarly, in the second period we have:
\begin{gather}
\left[
\begin{array}{ccc}
\delta a_1(\tau)\\b_1(\tau)\\b_2^{\dag}(\tau)
\end{array}\right]
=\left[
\begin{array}{ccc}
\cosh(r\Delta\tau_1)&0&i\sinh(r\Delta\tau_1)\\0&1&0\\-i\sinh(r\Delta\tau_1)&0&\cosh(r\Delta\tau_1)
\end{array}\right]
\left[\begin{array}{ccc}
\delta a_1(\tau_1)\\b_1(\tau_1)\\b_2^{\dag}(\tau_1)
\end{array}\right],
\end{gather}
where $r=\bar{a}_2g_2/\bar{a}_1g_1$ is the ratio of the coupling strength between the blue-detuned and red-detuned type interactions, and $\Delta\tau_1=\tau-\tau_1$. As for the third period:
\begin{gather}
\left[
\begin{array}{ccc}
\delta a_1(\tau)\\b_1(\tau)\\b_2^{\dag}(\tau)
\end{array}\right]
=\left[
\begin{array}{ccc}
\cos(\Delta\tau_2)&i\sin(\Delta\tau_2)&0\\i\sin(\Delta\tau_2)&\cos(\Delta\tau_2)&0\\0&0&1
\end{array}\right]
\left[\begin{array}{ccc}
\delta a_1(\tau_2)\\b_1(\tau_2)\\b_2^{\dag}(\tau_2)
\end{array}\right],
\end{gather}
where $\Delta\tau_2=\tau-\tau_2$. Now we can get the final state of the system at $\tau=\tau_1+\tau_2$
\begin{gather}
\left[
\begin{array}{ccc}
\delta a_1(\tau_1+\tau_2)\\b_1(\tau_1+\tau_2)\\b_2^{\dag}(\tau_1+\tau_2)
\end{array}\right]
=
M_1M_2M_1
\left[
\begin{array}{ccc}
\delta a_1(0)\\b_1(0)\\b_2^{\dag}(0)
\end{array}\right],\label{eq13}\\
\notag\\
M_1=
\left[
\begin{array}{ccc}
\cos(\tau_1)&i\sin(\tau_1)&0\\i\sin(\tau_1)&\cos(\tau_1)&0\\0&0&1
\end{array}\right],\\
\notag\\
M_2=
\left[
\begin{array}{ccc}
\cosh[r(\tau_2-\tau_1)]&0&i\sinh[r(\tau_2-\tau_1)]\\0&1&0\\-i\sinh[r(\tau_2-\tau_1)&0&\cosh[r(\tau_2-\tau_1)]
\end{array}\right].\\
\notag
\end{gather}
If $\cos(\tau_1)=0$, at $\tau=\tau_1+\tau_2$ the evolution matrix in Eq.(\ref{eq13}) becomes:
\begin{gather}
M_1M_2M_1=
\left[
\begin{array}{ccc}
-1&0&0\\0&-\cosh[r(\tau_2-\tau_1)]&-\sinh[r(\tau_2-\tau_1)]\\0&\sinh[r(\tau_2-\tau_1)]&\cosh[r(\tau_2-\tau_1)]
\end{array}
\right].
\label{eq14}
\end{gather}
Eq.(\ref{eq14}) shows at the instant $\tau=\tau_1+\tau_2$, the optical mode decouples with the two $LC$ modes, and the two $LC$ modes form the Bogoliubov modes. Moreover, in the case of initial vacuum states, the Bogoliubov modes turn to two-mode squeezed vacuum state \cite{PhysRevA.88.043802}.

\subsection{Parallel scheme}
In the parallel scheme, we want to set LC1 in the red-detuned regime, meanwhile LC2 is in the blue-detuned regime. Therefore, we need to apply two driving lasers of suitable frequencies simultaneously. The total Hamiltonian can be expressed as:
 \begin{gather}
 H=H_0+H_I+H_d',\\
 H_0=\hbar\omega_{a1}a_1^{\dag}a_1+\sum_{i=1}^{2}{\hbar\omega_{bi}b_i^{\dag}b_i},\\
 H_I=-\sum_{j=1}^{2}{\hbar g_j a_1^{\dag}a_1(b_j+b_j^{\dag})},\\
 H_d'=\hbar \sum_{j=1}^{2}(-1)^{j}E_j(a_1^{\dag}e^{-i\omega_{Lj}t}+a_1e^{i\omega_{Lj}t}).
 \end{gather}
 Here we apply driving signals with real amplitudes $E_j(j=1,~2)$ and initial phases $\phi_1=-\frac{\pi}{2},~\phi_2=\frac{\pi}{2}$.

 Then we can use similar approach as used in \cite{PhysRevA.86.012318} to simplify the Hamiltonian of our system. In the interaction picture, the total Hamiltonian becomes:
\begin{gather}
H'=-\sum_{j=1}^{2}{\hbar g_j a_1^{\dag}a_1(b_je^{-i\omega_{bj}t}+b_j^{\dag}e^{i\omega_{bj}t})}\notag\\
  +\hbar \sum_{j=1}^{2}(-1)^{j}E_j\left[a_1^{\dag}e^{(-1)^{j-1}i\Delta_jt}+a_1 e^{(-1)^j i\Delta_jt}\right],~\label{eq15}
\end{gather}
where $\Delta_1=\omega_{a1}-\omega_{L1}$, $\Delta_2=\omega_{L2}-\omega_{a1}$.

To engineer our desired coupling, we need an unitary transformation with the following defined unitary operator:
\begin{gather}
U=T\exp\left\{-i\int_0^t{\sum_{j=1}^{2}(-1)^{j}E_j\left[a_1^{\dag}e^{(-1)^{j-1}i\Delta_jt}+a_1 e^{(-1)^j i\Delta_jt}\right]}d t\right\},
\end{gather}
where $T$ is the time-ordering operator. Unlike the cascade scheme requiring extreme strong lasers, the parallel scheme requires lasers with relatively weak intensities: $E_j/\Delta_j<<1$ such that we can keep the leading term of $E_j/\Delta_j$. Eq.(\ref{eq15}) then becomes,
\begin{gather}
\begin{align}
H^P&=U^{\dag}\left(H'-i\hbar\frac{\partial}{\partial t}\right)U\notag\\
&=-\sum_{k=1}^{2}\hbar g_k \Big\{a_1^{\dag}a_1+\sum_{j=1}^2\frac{E_j}{\Delta_j}
\left[a^{\dag}(e^{(-1)^{j+1}i\Delta_jt}-1)+\textmd{H.c.}\right]\Big\}\notag\\
&\times(b_ke^{-i\omega_{bk}t}+b_k^{\dag}e^{i\omega_{bk}t}).
\end{align}
\end{gather}
Through setting $\Delta_1=\omega_{b1},\Delta_2=\omega_{b2}$ and using the rotating-wave approximation, $H^P$ turns to:
\begin{gather}
H^P_{eff}=-\hbar g_1\frac{E_1}{\Delta_1}(a_1^{\dag}b_1+a_1b_1^{\dag})-\hbar g_2 \frac{E_2}{\Delta_2}(a_1^{\dag}b_2^{\dag}+a_1b_2).~\label{eqnew1}
\end{gather}
Eq.(\ref{eqnew1}) yields the Langevin equation of this system,
\begin{gather}
\frac{d}{d\tau}
\left[
\begin{array}{ccc}
a_1\\b_1\\b_2^{\dag}
\end{array}
\right]
=\left[
\begin{array}{ccc}
-k_0&i&ir\\i&-k_1&0\\-ir&0&-k_2
\end{array}
\right]
\left[
\begin{array}{ccc}
a_1\\b_1\\b_2^{\dag}
\end{array}
\right]
+\left[
\begin{array}{ccc}
f_0\\f_1\\f_2^{\dag}
\end{array}
\right],~\label{eq17}
\end{gather}
where $\tau$ and $r$ are now defined by $\tau=g_1\frac{E_1}{\Delta_1}t$, $r=\frac{E_2\Delta_1}{E_1\Delta_2}$. $\{k_i,i=0,1,2\}$ are non-dimensional decay rates defined by $k_i=\frac{\Gamma_i\Delta_1}{2E_1g_1}$ and $\{\Gamma_i,i=1,2,3\}$ are those decay rates of each mode in Eq.(\ref{eq17}); $f_i=\frac{F_i\Delta_1}{g_1E_1}$ with $\{F_i,i=0,1,2\}$ the noise operators of each modes. According to \cite{scully1997quantum}, $\langle F_i^{\dag}(t)F_j(t')\rangle_R=\Gamma_i n_{i,th}\delta_{ij}\delta(t-t')$, it is straightforward that:
\begin{gather}
\langle f_i^{\dag}(\tau)f_j(\tau')\rangle_R=2k_i n_{i,th}\delta_{ij}\delta(\tau-\tau').~\label{eq18}
\end{gather}
We solve the Heisenberg equation and Langevin equation for the non-dissipative and dissipative cases, respectively.

(i)If $k_i=0,i=0,1,2$, Eq.(\ref{eq17}) converts into a homogeneous equation. The time evolution of the system is:
\begin{gather}
\left[
\begin{array}{ccc}
a_1(\tau)\\b_1(\tau)\\b_2^{\dag}(\tau)
\end{array}
\right]
=M^P
\left[
\begin{array}{ccc}
a_1(0)\\b_1(0)\\b_2^{\dag}(0)
\end{array}
\right],
\end{gather}
\begin{gather}
M^P=
\left[
\begin{array}{ccc}
\cos(\sqrt{1-r^2}\tau)&i\frac{\sin(\sqrt{1-r^2}\tau)}{\sqrt{1-r^2}}&ir\frac{\sin(\sqrt{1-r^2}\tau)}{\sqrt{1-r^2}}\\
i\frac{\sin(\sqrt{1-r^2}\tau)}{\sqrt{1-r^2}}&\frac{\cos(\sqrt{1-r^2}\tau)-r^2}{1-r^2}&\frac{(\cos(\sqrt{1-r^2}\tau)-1)r}{1-r^2}\\
-ir\frac{\sin(\sqrt{1-r^2}\tau)}{\sqrt{1-r^2}}&\frac{(1-\cos(\sqrt{1-r^2}\tau))r}{1-r^2}&\frac{1-r^2\cos(\sqrt{1-r^2}\tau)}{1-r^2}
\end{array}
\right],
~\label{eq20}
\end{gather}

(ii) If $k_i\neq0,i=0,1,2$, the time evolution of this system can be solved using the theory of linear differential equations.
We assume that:
\begin{gather}
c=
\left[
\begin{array}{ccc}
-k_0&i&ir\\i&-k_1&0\\-ir&0&-k_2
\end{array}
\right].
\end{gather}
The eigenvalues and corresponding column eigenvectors of matrix $c$ are $\{\lambda_i,\vec{u}_i,~i=1,2,3\}$, and the time evolution of such system can be written as:
\begin{gather}
\left[
\begin{array}{ccc}
a_1(\tau)\\b_1(\tau)\\b_2^{\dag}(\tau)
\end{array}
\right]=
X(\tau)
\left[
\begin{array}{ccc}
a_1(0)\\b_1(0)\\b_2^{\dag}(0)
\end{array}
\right]\notag\\
+\int_0^{\tau}X(\tau-\tau')
\left[
\begin{array}{ccc}
f_0(\tau')\\f_1(\tau')\\f_2^{\dagger}(\tau')
\end{array}
\right]d\tau',\\
\notag\\
X(\tau)=[e^{\lambda_1\tau}\vec{u}_1,e^{\lambda_2\tau}\vec{u}_2,e^{\lambda_3\tau}\vec{u}_3]
[\vec{u}_1,\vec{u}_2,\vec{u}_3]^{-1}.
\end{gather}

It is not necessary for us to get exact analytic solutions in this case, so we can get numerical solutions from the expressions above.

\begin{figure}[!ht]
  \centering
  \includegraphics[width=8cm]{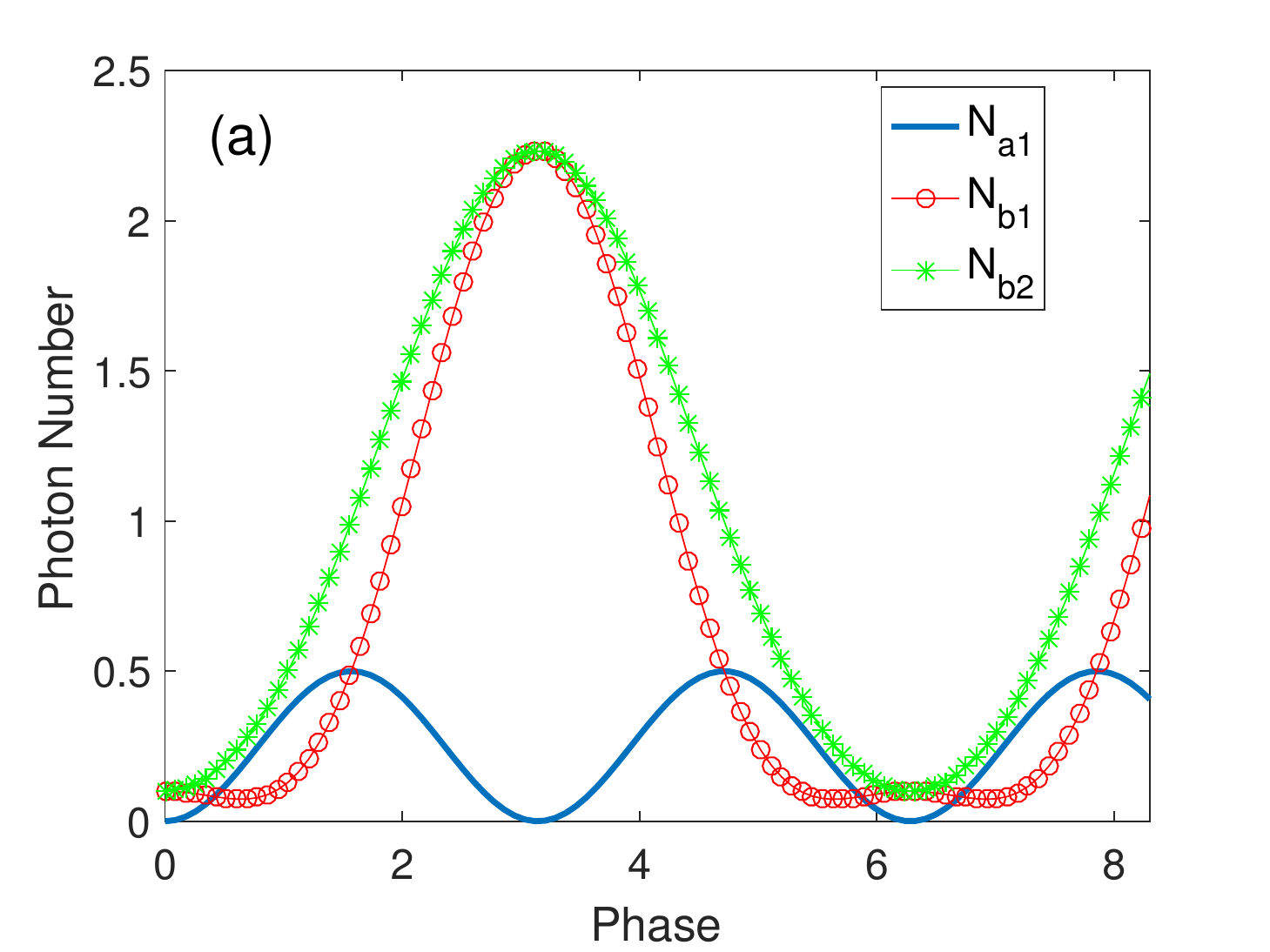}\\
  \includegraphics[width=8cm]{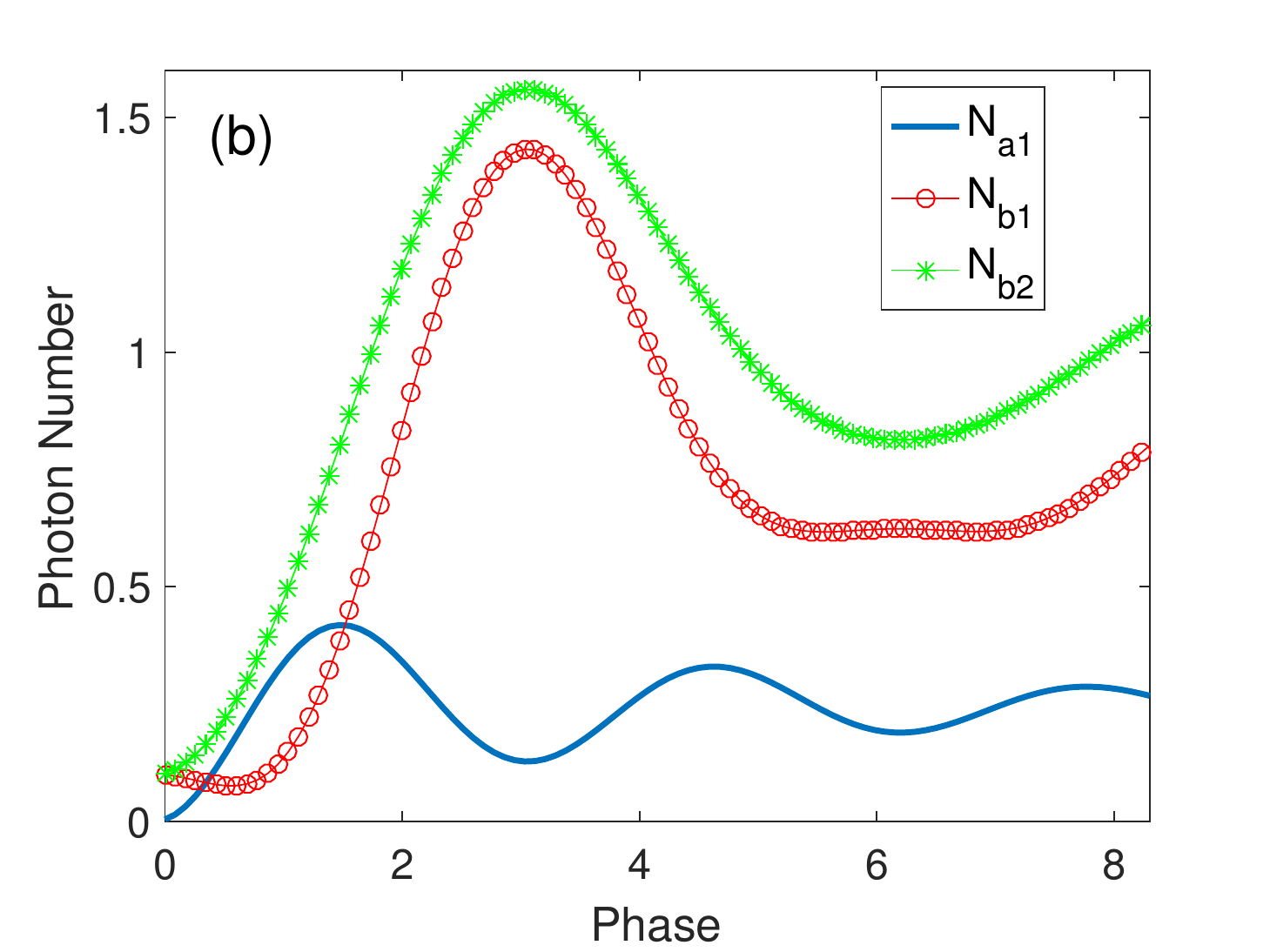}
  \caption{The time evolution of fluctuating photon number for different values of the non-dimensional decay rate $k_i$. $N_{a1}$, $N_{b1}$ and $N_{b2}$ are photon numbers for optical cavity, LC1 and LC2, respectively. (a) $r=0.5,~k_0=k_1=k_2=0$, (b) $r=0.5,k_0=k_1=k_2=0.1$. The initial condition for both (a) and (b) is $N_{a1}(0)=n_{0,th}=0,~N_{b1}(0)=N_{b2}(0)=n_{1,th}=n_{2,th}=0.1$. In (a) the ideal case, at each instant $n\pi$, the photon number of optical mode goes to zero and those numbers of LC1 and LC2 become equal, which is in accordance with the decoupling of optical mode and the entanglement of two superconducting microwave resonators. From (b), we can find that even for optical cavity, the photon number can not be zero at steady state.~\label{fg2}}
\end{figure}

As for the non-dissipative case, when $\sqrt{1-r^2}\tau=\pi$, Eq.(\ref{eq20}) becomes:
\begin{gather}
M^P=
\left[
\begin{array}{ccc}
-1&0&0\\
0&-\frac{1+r^2}{1-r^2}&-\frac{2r}{1-r^2}\\
0&\frac{2r}{1-r^2}&\frac{1+r^2}{1-r^2}
\end{array}
\right].
~\label{eq21}
\end{gather}
 Eq.(\ref{eq21}) indicates that at the instant $T_{\pi}=\pi/\sqrt{1-r^2}$, the optical mode decouples from the dynamics of the system. We assume $\cosh(\xi)=(1+r^2)/(1-r^2),~\sinh(\xi)=2r/(1-r^2)$ and introduce the operator $S=e^{\xi[b_1(0)b_2(0)-b_1^{\dag}(0)b_2^{\dag}(0)]}$,
 then $b_1(T_{\pi})=-Sb_1(0)S^{\dag},~b_2^{\dag}(T_{\pi})=Sb_2^{\dag}(0)S^{\dag}$, indicating the two superconducting microwave resonators are prepared in the Bogoliubov modes. If the initial states of the superconducting microwave resonators are vacuum states, they will be prepared in the two-mode squeezed state with the squeezing parameter $\xi=\tanh^{-1}[2r/(1+r^2)]$  at the moment $T_{\pi}$. Fig.$\ref{fg2}$ shows the time evolution of the photon numbers. For the non-dissipative case shown in Fig.$\ref{fg2}$(a), at the instant $T_{\pi}$ the photon number of the optical cavity drops to 0 and the photon numbers of the superconducting microwave resonators become equal. This is in accordance with the conclusion that at that moment the optical mode decouples from the dynamics of the system and the microwave modes get entangled. From Fig.$\ref{fg2}$(b), we can see that the periodic fluctuations of photon numbers are impeded by the dissipations. As a result, the photon number of the optical cavity can't decrease to 0, reflecting that some photons of microwave modes still interact with the remaining photons of optical modes. Therefore, when the effect of dissipations is notable, there is no such instant as $T_{\pi}$ that the photons of the two microwave modes can be entangled completely
\graphicspath{{fig/}}

\begin{figure}[!ht]
\center
\includegraphics[width=8cm]{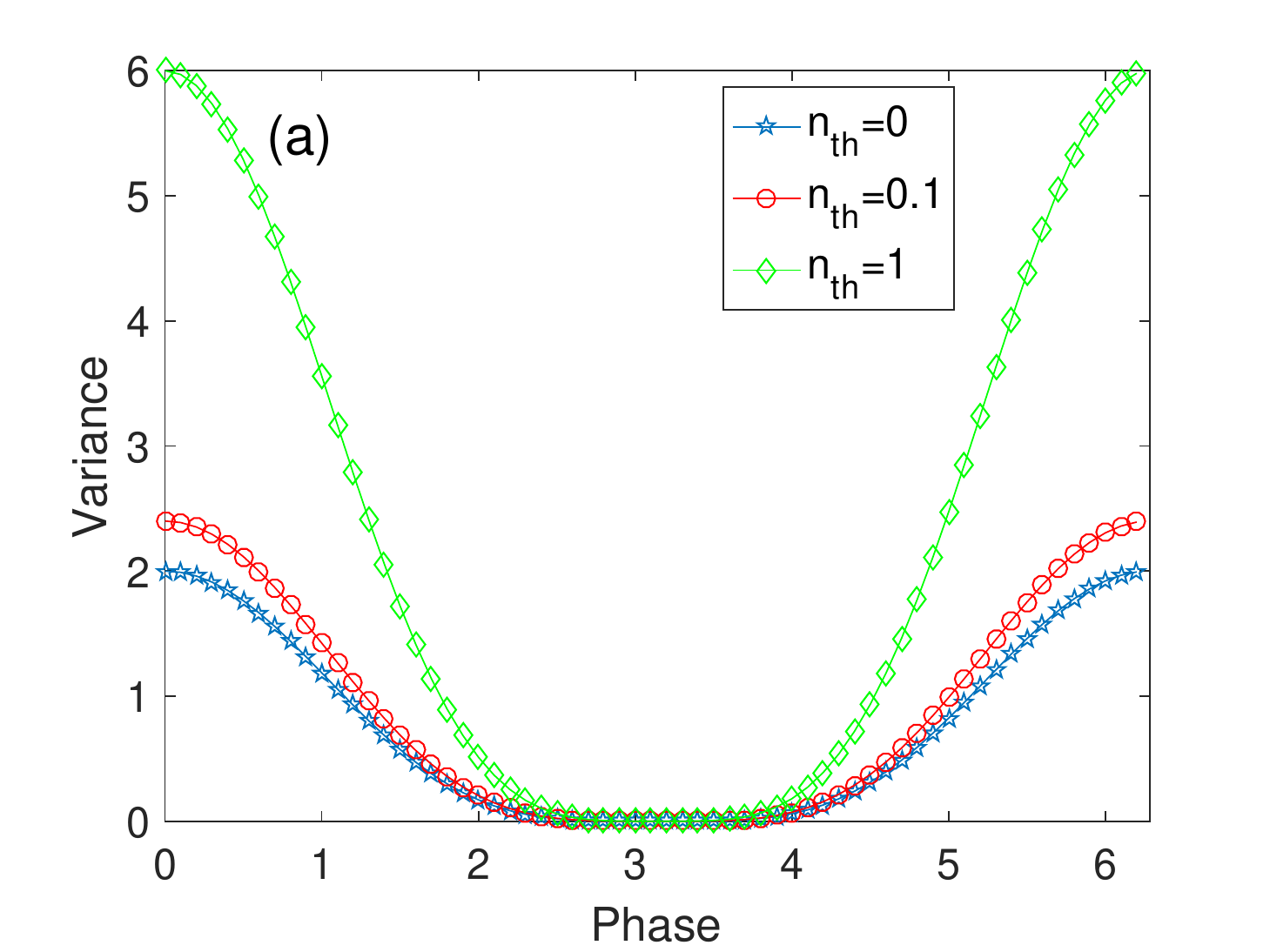}
\includegraphics[width=8cm]{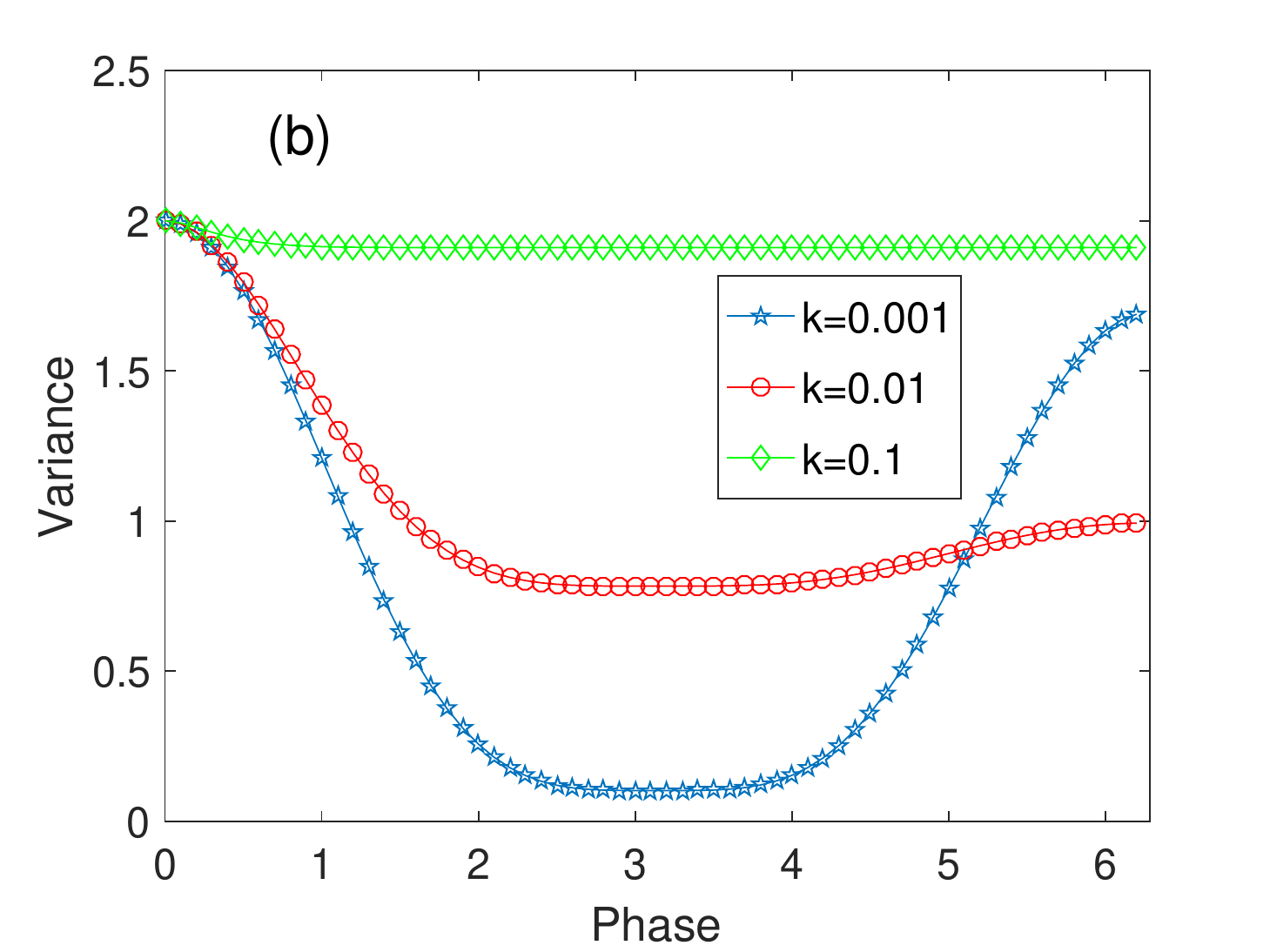}
\caption{The total variance V.S. phase for different values of parameters: (a) without dissipation, $r=1-10^{-3},~n_{1,th}=n_{2,th}=n_{th}$, and $n_{th}=0,~0.1,~1$; (b) with dissipation, $r=1-10^{-3}$, $n_{th}=0$, $k_0=k_1=k_2=k$, and $k=0.001,~0.01,~0.1$.~\label{fg3}}
\end{figure}

In order to investigate the entanglement of modes $b_1$ and $b_2$, we need the total variance $V=\langle (\Delta u)^2+(\Delta v)^2\rangle$ of EPR-like operators $u=x_1+x_2,~v=p_1-p_2$, with $x_i=(b_i+b_i^{\dag})/\sqrt{2}$ and $p_i=-i(b_i-b_i^{\dag})/\sqrt{2}$, $i=1,2$ \cite{PhysRevA.88.043802}. The two-mode Gaussian state is entangled if and only if $V<2$ \cite{PhysRevA.88.043802,PhysRevLett.84.2722}. Especially, for the two-mode squeezed vacuum state, $V=2e^{-2\xi}$. We explore the effects of the dissipation and the initial thermal conditions to the entanglement of this system as illustrated in Fig.$\ref{fg3}$. In Fig.$\ref{fg3}$(a), we change the initial thermal condition of each mode, and find that at the neighborhood of $T_{\pi}$ the differences of all curves disappear. We've already known that at the instant $T_{\pi}$, the two microwave modes are entangled. Therefore, the entanglement of this system is insensitive to the initial thermal conditions. But in Fig.$\ref{fg3}$ (b), when we modulate the decay rates of all modes, the total variances vary greatly with them. Thus, the low-decay condition should be satisfied in order to get better entanglements.
\graphicspath{{fig/}}

\subsection{Dissipative dynamical scheme}

\begin{figure}[!ht]
  \centering
  \includegraphics[width=8cm]{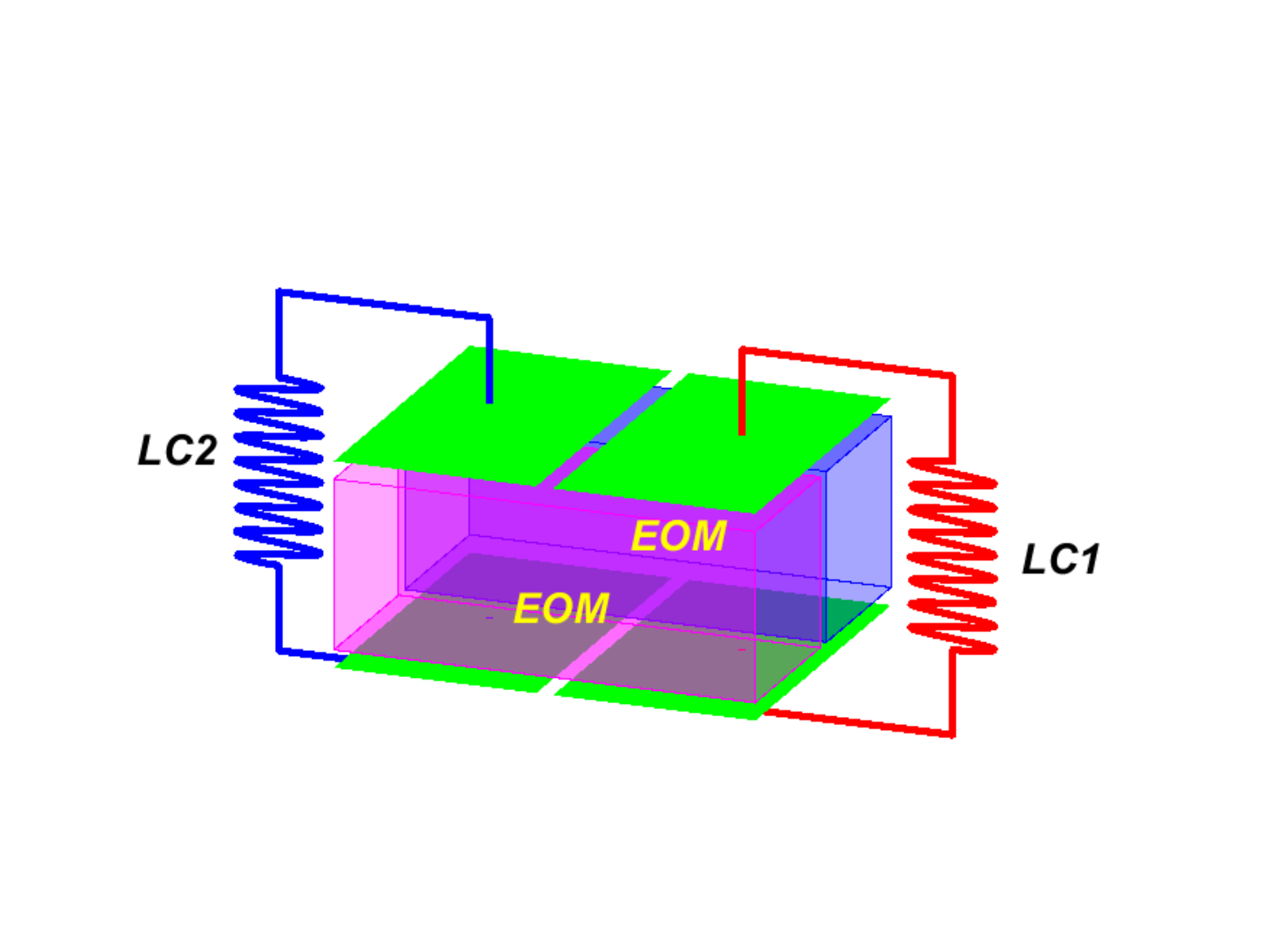}\\
  \caption{The setup for the dissipative dynamical scheme. Two optical cavities are filled by electro-optic media and modulated by both LC1 and LC2 via the electro-optic effect. LC1 is in the red-detuned regime with respect to the previous optical cavity and blue-detuned regime to the second one, but LC2 is just on the contrary.
~\label{fg4}
}
\end{figure}

The system in the parallel scheme is very sensitive to dissipations as shown in Fig.\ref{fg3}, and so does the cascaded scheme. This indicates in previous schemes, we need to attenuate the dissipation as much as possible. Thus, in many cases, high-Q cavities or resonators are necessary. But we can also prepare our target states with low-Q optical cavities("bad cavities"). In the dissipative dynamical scheme, large decay rates of the optical modes are required due to the fact that the dissipative effects of the optical modes here are treated as a useful resources. Such schemes in other hybrid quantum systems have been explored previously\cite{PhysRevA.88.043802}. However, in our system, what we use is the optical thermal noise, where the mean photon number at thermal equilibrium is $n_{0,th}\approx0$. Therefore, through the electro-optic system we can get more ideal two-mode squeezed vacuum states at the same temperature, compared with the optomechanical systems. To realize our scheme, it is necessary to put LC1 and LC2 in both red- and -blue detuned regimes at the same time. One approach is to add another optical cavity of frequency $\omega_{a2}$ satisfying $|\omega_{a2}/\omega_{a1}-1|<<1$ paralleled to the previous one as shown in Fig.\ref{fg4}. Through modulating the parameters in Eq.(\ref{eq5}), the coupling strengths of the "beam-splitter and "two-mode squeezing" interactions in the second optical cavity can keep the same as the first one.
\graphicspath{{fig/}}
The ideal situation for this scheme is $k_0>>1>>k_1$, where $k_0$ stands for the non-dimensional decay rate of two optical cavities while $k_1$ denotes the non-dimensional decay rate of two microwave resonators. Therefore, we can ignore the dissipation of two microwave modes, Then the Langevin equation of this system becomes:
\begin{gather}
\frac{d}{d\tau}
\left[
\begin{matrix}
a_1(\tau)\\a_2^{\dag}(\tau)\\b_1(\tau)\\b_2^{\dag}(\tau)
\end{matrix}
\right]=
\left[
\begin{matrix}
-k_0&0&i&ir\\0&-k_0&-ir&-r\\i&ir&0&0\\-ir&-i&0&0
\end{matrix}
\right]
\left[
\begin{matrix}
a_1(\tau)\\a_2^{\dag}(\tau)\\b_1(\tau)\\b_2^{\dag}(\tau)
\end{matrix}
\right]
+\left[
\begin{matrix}
f_{a1}(\tau)\\f_{a2}^{\dag}(\tau)\\0\\0
\end{matrix}
\right].~\label{eq25}
\end{gather}
Here the definition of all the non-dimensional variables has the same form as that in the parallel scheme. In the ideal situation, we can make the adiabatic approximations to the optical modes:
\begin{gather}
a_1(\tau)=\frac{i}{k_0}(b_1(\tau)+rb_2^{\dag}(\tau))+\frac{f_{a1}(\tau)}{k_0},~\label{eq28}\\
a_2^{\dag}(\tau)=-\frac{i}{k_0}(rb_1(\tau)+b_2^{\dag}(\tau))+\frac{f_{a2}^{\dag}(\tau)}{k_0}.~\label{eq29}
\end{gather}
Inserting Eq.(\ref{eq28}) and Eq.(\ref{eq29}) into Eq.(\ref{eq25}) yields the relationship between $b_1$ and $b_2^{\dag}$:
\begin{gather}
\frac{d}{d\tau}
\left[
\begin{matrix}
b_1(\tau)\\b_2^{\dag}(\tau)
\end{matrix}
\right]=
\frac{1}{k_0}
\left[
\begin{matrix}
r^2-1&0\\0&r^2-1
\end{matrix}
\right]
\left[
\begin{matrix}
b_1(\tau)\\b_2^{\dag}(\tau)
\end{matrix}
\right]\notag\\+
\frac{i}{k_0}
\left[
\begin{matrix}
f_{a1}(\tau)+rf_{a2}^{\dag}(\tau)\\-rf_{a1}(\tau)-f_{a2}^{\dag}(\tau)
\end{matrix}
\right].
\label{eq30}
\end{gather}

We introduce some new variables and operators to simplify our expressions.
\begin{gather}
x=\frac{(1-r^2)\tau}{k_0},~x'=\frac{(1-r^2)\tau'}{k_0},~\label{eq31}\\
\tilde{f}_{ai}(x)=\frac{f_{ai}(x)}{\sqrt{2k_0}},~i=1,2,~\label{eq32}\\
\tilde{S}(x,\varsigma)=\exp\left[\varsigma(\tilde{f}_{a1}(x)\tilde{f}_{a2}(x)-\tilde{f}_{a1}^{\dag}(x)\tilde{f}_{a2}^{\dag}(x))\right],~\label{eq33}
\end{gather}
where $\varsigma=\arctan\left[r\right]$ is the squeezing parameter of optical modes, and by the definition of $\tilde{f}_{ai}$, we can get $\langle\tilde{f}_{ai}(x)\tilde{f}_{aj}^{\dag}(x')\rangle_R=\delta_{ij}\delta(x-x')$. With Eq.(\ref{eq31})-(\ref{eq33}), the solution of Eq.(\ref{eq30}) can be expressed as:
\begin{gather}
b_1(x)=e^{-x}b_1(0)+\sqrt{2}i\int_0^xe^{-(x-x')}\tilde{S}(x')\tilde{f}_{a1}(x')\tilde{S}^{\dag}(x')dx',~\label{eq34}\\
b_2^{\dag}(x)=e^{-x}b_2^{\dag}(0)-\sqrt{2}i\int_0^xe^{-(x-x')}\tilde{S}(x')\tilde{f}_{a2}^{\dag}(x')\tilde{S}^{\dag}(x')dx'.~\label{eq35}
\end{gather}
From Eq.(\ref{eq31})-(\ref{eq35}), we can find the stable condition for this system is $r<1$, under which as $x\rightarrow\infty$ the system will converge to the final state of the Bogoliubov modes composed of optical modes. As in the parallel scheme, we calculate the total variance of EPR-like operators composed of $b_1$ and $b_2^{\dag}$ in the stable case, $V=\lim\limits_{x\to\infty}{\frac{2(1-r)}{1+r}(1-e^{-2x})=\frac{2(1-r)}{1+r}}$. That is exactly the total variance of the ideal two-mode vacuum state $2e^{-2\varsigma}$ with squeezing parameter $\varsigma=\arctan[r]$. Thus, under the assumption that $k_0\gg1\gg k_1$, no mater what the initial condition is, the two microwave modes will finally evolve to the two-mode squeezed vacuum state, definitely.

\begin{figure}[!ht]
  \centering
  \includegraphics[width=8cm]{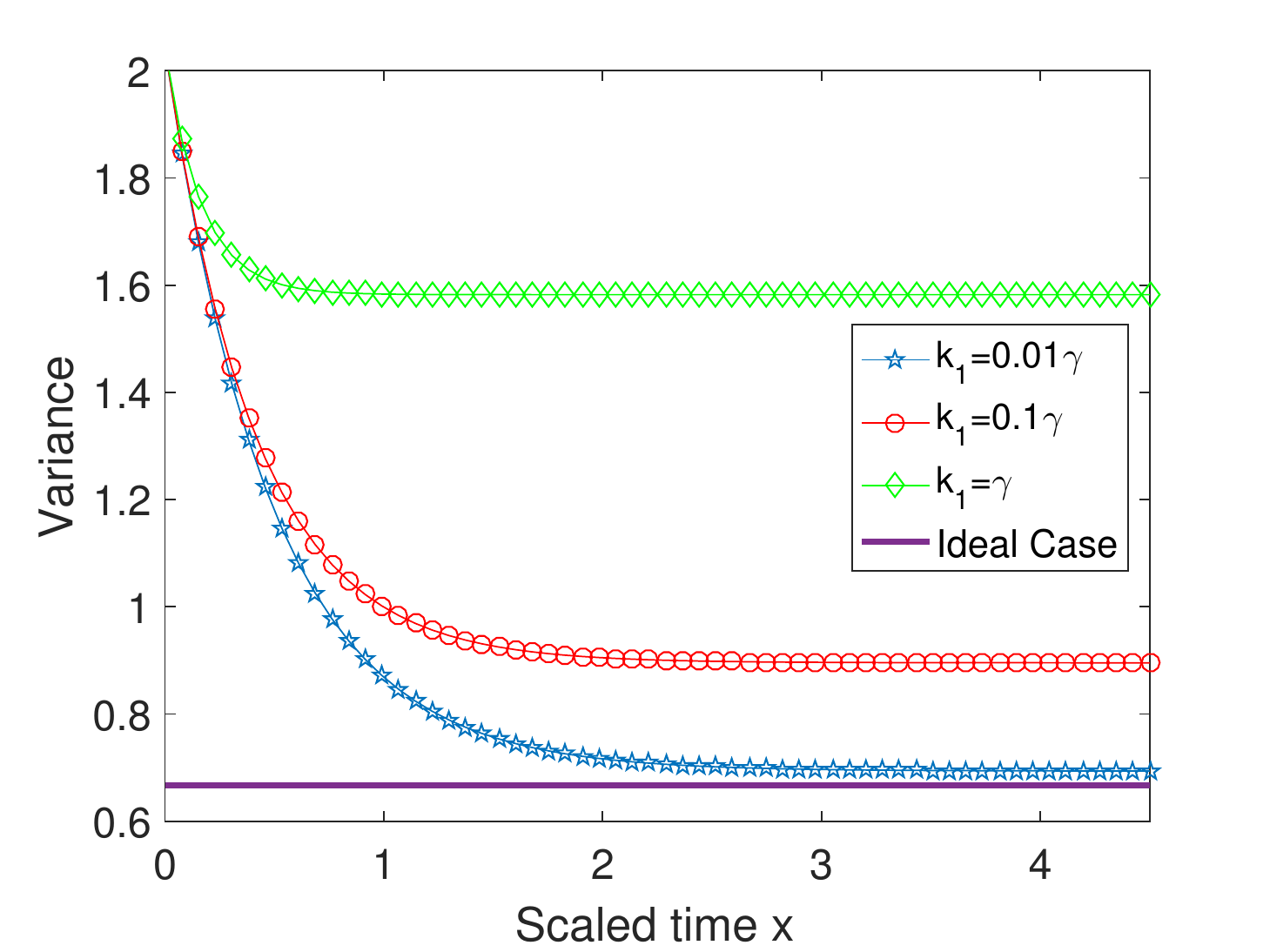}\\
  \caption{Plot of the total variance $V$ V.S. scaled time $x$ under different decay conditions, with parameters $r=0.5$, $n_{b,th}=0.01$, $k_0=10$, together with the result for an ideal two-mode squeezed state.  The effective decay rate of the subsystem formed by two superconducting microwave resonators in Eq.(\ref{eq30}) is $\gamma=\frac{2(1-r^2)}{k_0}=0.15$.~\label{fg5}}
\end{figure}

Let's consider this scheme in a more general case, where we only make adiabatic approximations to optical modes. Then such total variance becomes $V=\Sigma_{i=1}^2 e^{-2(x+k_i\tau)}(2n_{th,i}+1)+\frac{(1-r)^2+(n_{th,i}+1)k_0k_i}{1-r^2+k_0k_i}\times[1-e^{-2(x+k_i\tau)}]$. The effective decay rate in Eq.(\ref{eq30}) varies inversely with the decay rate of optical modes, and that explains why we need the optical cavity with a large decay rate. In Fig.$\ref{fg5}$, we present the time evolution of the total variance under different decay rates of microwave modes as well as the result of an ideal two-mode squeezed vacuum state. The initial conditions are chosen as the ground states for the optical cavities and the thermal states for the two microwave modes. From this figure, we know that when the decay rates of microwave modes are much smaller than the effective decay rate, at steady state we can prepare nearly ideal two-mode vacuum squeezed states.
\graphicspath{{fig/}}

\section{Experimental Feasibility}\label{sec1}
We now talk about the experimental parameters. For the cascaded scheme, it is feasible to take the capacitances and inductances of $LC$ resonators as $C_1=C_2=$40fF, $L_1=$70nH, $L_2=$25nH. Similar to that electro-optic system reported by Mankei Tsang\cite{PhysRevA.81.063837}, we can take the electro-optic coefficient $n^3r_0\approx$300pm/V, the resonance frequency of optical cavity $\omega_{a1}\approx2\pi\times200$THz, the decay rate of superconducting microwave resonators $\Gamma_i\approx2\pi\times1$KHz, $i=1,~2$, and assume the distance between two planes of each capacitor $d\approx10\mu$m. The pump power is able to reach $P=10$mW\cite{Ilchenko:03}. Thus, the coefficients $g_i$ given by Eq.(\ref{eq5}) can reach $g_1\approx 2\pi \times 15$KHz, $g_2\approx2\pi\times19$KHz, and $\omega_{b1}\approx2\pi\times 3$GHz, $\omega_{b2}\approx2\pi\times5$GHz. In the "overcoupled" case, we can also work out the mean photon number of optical cavity caused by external pump $\bar{n}_{cav,i}$ through the following equation\cite{RevModPhys.86.1391}
\begin{gather}
	\bar{n}_{cav,i}=\frac{\Gamma}{\Delta_i^2+\left(\Gamma/2\right)^2}\frac{P}{\hbar\omega_{a1}},
\end{gather}
where $\Gamma$ is the total loss rate of optical cavity, which is dominated by external loss rate of the cavity. For the Q-factor of optical cavities can reach $10^8$, we have $\Gamma_0\approx2\pi\times 0.3$MHz. In our case we choose $\Delta_i=\omega_{bi}$, and get $\bar{n}_{cav,1}\approx400,\bar{n}_{cav,2}\approx144$. Then the effective electro-optic coupling strength can reach $\sqrt{\bar{n}_{cav,1}}g_1\approx2\pi\times0.3$MHz, $\sqrt{\bar{n}_{cav,2}}g_2\approx2\pi\times0.23$MHz. If we set the time for "two-mode squeezed" interaction is $T_2\sim1.6\mu$s, the operation time for generating target states will be $T_c=\pi/\left(\bar{a}_1g_1\right)+T_2\sim3.2\mu$s in the cascaded scheme.

As for parallel and dissipative dynamical schemes, we assume the pump power is relatively low, i.e. $P=10\mu$W, the distance $d=5\mu$m, the capacitances and inductances of superconducting microwave resonators $C_1=C_2=4$fF, $L_1=700$nH, $L_2=250$nH, $\Gamma_1=\Gamma_2\approx2\pi\times 1$KHz, and use the optical cavity with resonance frequency $2\pi\times 1500$THz$(\lambda\approx200nm)$ to ensure the validity of the expansion applied in both schemes. In our case, $\Gamma
\ll\Delta_i$. Then the amplitudes of driving lasers in both schemes can be expressed as $E_i=\sqrt{\bar{n}_{cav,inew}}\omega_{bi},~i=1,~2$. Therefore, the operation time for the parallel scheme to generate target states is $T_p=\pi/\left[g_1\sqrt{\bar{n}_{cav,1new}-\bar{n}_{cav,2new}}\right]\sim3.8\mu$s, and the time for reaching stationary state of the dissipative dynamical scheme $T_d=\Gamma/\left[2g_1^2\left(\bar{n}_{cav,1new}-\bar{n}_{cav,2new}\right)\right]\sim1.3\mu$s. These times are much shorter than the photon lifetime in superconducting microwave resonators. 

Further more, if it is allowed to realize large inductance, the effective electro-optic coupling strength will exceed $2\pi\times 1$MHz. For example, we take $L=63\mu$H and keep other parameters the same as those in the cascaded scheme, its effective electro-optic coupling strength will reach $2\pi\times$1.6MHz and the optical loss rate can be ignored for simplicity.

\begin{figure}[!ht]
	\centering
	\includegraphics[width=8cm]{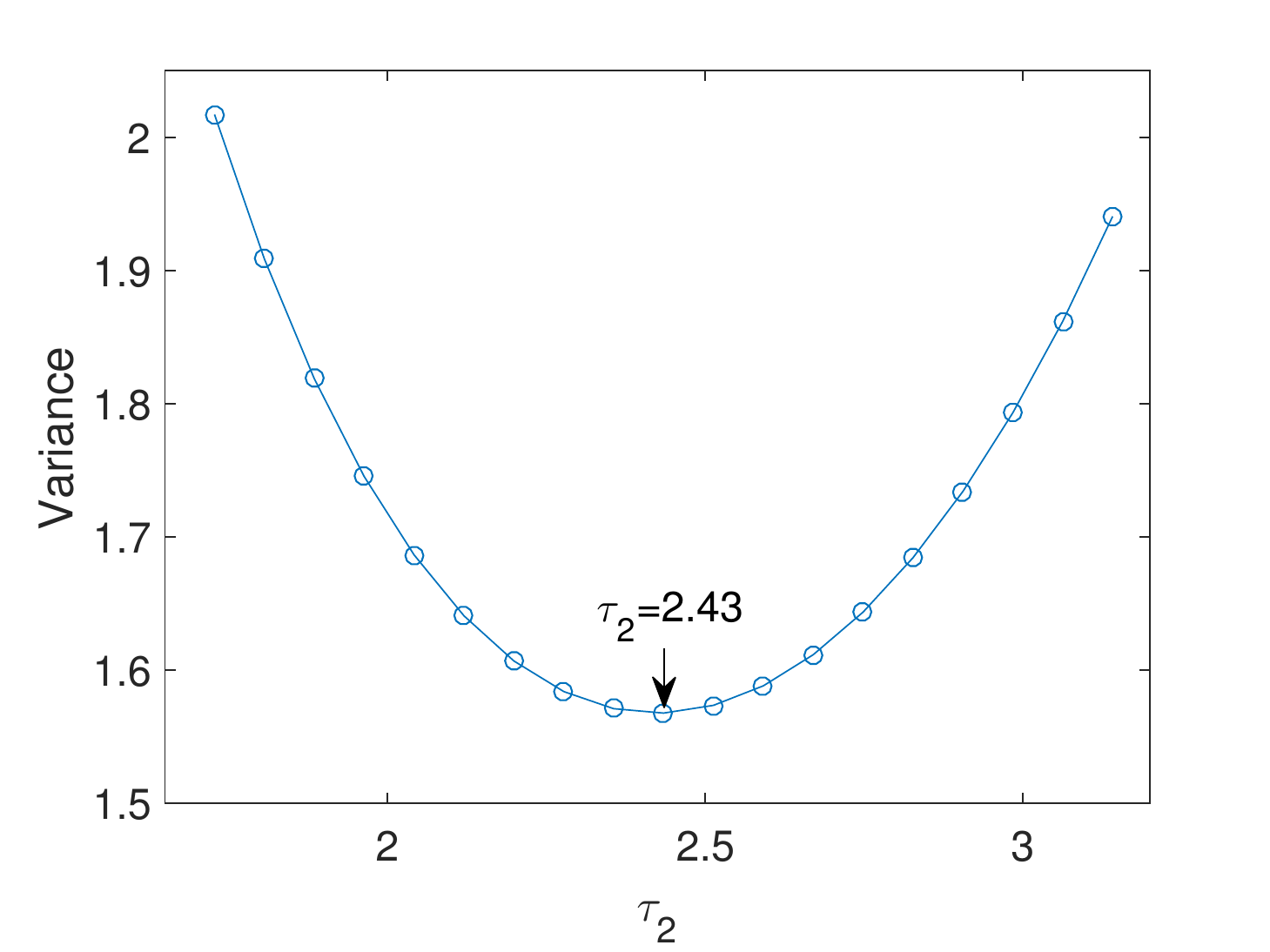}\\
	\caption{Plot of the total variance $V$ V.S. scaled time $\tau_2$ under above experimental parameters. And we can see that the optimized scaled time is $\tau_2=2.43$, or equally $T_2\approx1.3\mu$s.~\label{fg6}}
\end{figure}

We are also concerned about the entanglement properties of systems in each scheme. From Eq.(\ref{eq14}), we can see the squeezed parameter of the cascaded scheme in ideal case is determined by $\zeta=r\left(\tau_2-\tau_1\right)$. Obviously, $\zeta$ will increase with the increase of $\tau_2$. But effects of dissipation and decoherence also grow when $\tau_2$ increases. Therefore, people need to strike a balance between both aspects. We assume the temperature is approximately 100mK. In the previous experimental parameters in the cascaded scheme, the thermal photon numbers are $n_{th,1}\approx0.3,~ n_{th,2}\approx0.1$. Through numerical simulation shown in Fig.\ref{fg6}, we can find the optimized scaled time $\tau_2$ is 2.43 or equally $T_2\approx1.3\mu$s, and the minimum of total variance is approximately 1.56. We find that the total variance is insensitive to the environment temperature when it is below 1K, but greatly relies on scaled decay rates of all modes $k_i=\Gamma_i/\left(2\bar{a}_1g_1\right),~i=0,~1,~2$. Thus, we can reset those related parameters to improve the quality of target states. For example, when we change capacities and inductances to $C_1=C_2=1$fF, $L_1=360$nH, $L_2=350$nH, at the same temperature, the minimal variance drops to $V\approx0.77$. 

In the parallel scheme, the total variance is greatly affected by not only scaled decay rates, but also the environment temperature. With those experimental parameters used for discussing operation times of this scheme, scaled decay rates are $k_0\approx0.9,~k_1\approx k_2\approx0.003$. At the temperature $T=100$mK, the lowest variance is 0.66, but at the temperature $T=1$K, the entanglement will be destroyed.

In order to discuss the total variance of the dissipative dynamical scheme, it is useful to simplify the expression of its stable variance as following:
\begin{gather}
	V=\frac{2\left(1-r\right)/\left(1+r\right)+2\alpha\left(n_{th,1}+n_{th,2}+2\right)}{1+2\alpha},\label{eq43}\\
	\alpha=k_i/\gamma,~i=1,2,\\
	\gamma=2\left(1-r^2\right)/k_0.\label{eq45}
\end{gather}
Where we've assumed two superconducting resonators have same the decay rate $\Gamma_1=\Gamma_2$, and $\gamma$ is the effective decay rate of the system. We choose $\Gamma_0=2\pi\times 30$MHz, $\Gamma_1=\Gamma_2\approx2\pi\times1.44$KHz and keep other parameters same as the parallel scheme. At the same temperature $T=100$mK, the total variance can decrease to 0.3.

\section{Conclusions}

To conclude, we propose three schemes to generate the entanglement of microwave photons with an electro-optic system, in which two superconducting microwave resonators are coupled by one or two optical cavities through electro-optic effect. The first two schemes are based on coherent control over the system to realize Bogoliubov modes consisting of two microwave modes while the last scheme is based on dissipative dynamics engineering, which exploits the thermal noises of two optical cavities as useful resources to entangle microwave modes. Compared to previous works, our electro-optic system can generate more ideal two-mode squeezed states in principle. These schemes based on the electro-optic system may have novel applications in quantum information processing.
\section*{Acknowledgments}
This work is supported by the NSFC under Grant
No. 11474227 and the Fundamental
Research Funds for the Central Universities.
\end{document}